# A Trust Model for Data Sharing in Smart Cities

Quyet H. Cao*‡, Imran Khan†, Reza Farahbakhsh‡, Giyyarpuram Madhusudan*,
Gyu Myoung Lee§, Noel Crespi‡

*Orange Labs, France, Email: {quyet.caohuu, giyyarpuram.madhusudan}@orange.com
†Schneider Electric Industries SAS, 38TEC, 38050, Grenoble Cedex 9, France, Email: imran@ieee.org
‡Institut Mines-Telecom, Telecom SudParis, CNRS UMR 5157, France,
Email: {reza.farahbakhsh, noel.crespi}@it-sudparis.eu
§Department of Computer Science, Liverpool John Moores University, UK, Email: g.m.lee@ljmu.ac.uk

*Abstract*—The data generated by the devices and existing infrastructure in the Internet of Things (IoT) should be shared among applications. However, data sharing in the IoT can only reach its full potential when multiple participants contribute their data, for example when people are able to use their smartphone sensors for this purpose. We believe that each step, from sensing the data to the actionable knowledge, requires trust-enabled mechanisms to facilitate data exchange, such as data perception trust, trustworthy data mining, and reasoning with trust related policies. The absence of trust could affect the acceptance of sharing data in smart cities. In this study, we focus on data usage transparency and accountability and propose a trust model for data sharing in smart cities, including system architecture for trust-based data sharing, data semantic and abstraction models, and a mechanism to enhance transparency and accountability for data usage. We apply semantic technology and defeasible reasoning with trust data usage policies. We built a prototype based on an air pollution monitoring use case and utilized it to evaluate the performance of our solution.

## I. INTRODUCTION

Data sharing in the Internet of Things (IoT) [1] in general and in the context of smart cities [2] in particular will only reach its full potential if data can be collected by multiple sources. One such example is that people are able to share their data related to different events by leveraging the sensing capabilities of their smartphones. This crowd-sensing is a recent trend [3] and may soon outperform traditional data collection methods such as using pre-installed sensors. However, crowd-sensing may involve privacy issues for device owners. For example, some of the data collected by smartphones may contain sensitive information such as the location data of the owners. In the context of smart cities, the data may come from a variety of sources, such as institutional actors, equipment manufacturers, network operators, infrastructure providers, service providers, and end users [4]. These data potentially undergo several transformations, such as aggregation and composition, before reaching their final destination. Another important aspect is that the IoT data may also be shared for common usage through linked data sets such as Linked Open Data [5]. Therefore successful, and in some cases meaningful data sharing in smart cities depends on the establishment of trusted relationships among participants. We believe that participants will share their data when they have the ability to control the use of their data.

To deal with this issue of trust and control, we have proposed a data usage control model to capture the diversity of obligations and constraints that data owners impose on the use of data [4]. However, the architectural support to provide data usage transparency and accountability is still lacking, motivating us to develop this type of architecture support for stakeholders in the context of shared platforms in smart cities. The stakeholders themselves can thus participate in the sequence of steps in the mechanism that enhances the transparency and accountability of data usage.

We use the concept of ontologies and introduce the notion of trust ontology, a formal representation of concepts related to data usage control requirements, to annotate the data generated by the devices or resources in smart cities. We have a semantic data model with which to present the number of entities, the states of these entities. This leads to increased flexibility in terms of data integration, modeling, and processing compared to our previous data model based on NGSI [6]. This approach is also aligned with the standardization reported in OneM2M [7] as it provides the required abstractions.

Moreover, we provide trust enforcement for shared data based on the consumers' requests and policies of data owners, allowing the IoT shared platform to keep track of data usage history. We then experiment further on a specific use case, using a logic reasoner [8] to provide tests based on defeasible reasoning. Trust-based Data Usage (TDU) is the name of our solution.

The main contributions of this paper are four-fold: (*i*) A multi-layer architecture for TDU - we describe a use case scenario, its background and main functional entities. We also include a semantic and abstraction discussion for data integration, modeling, and processing; (*ii*) A mechanism to enhance the transparency and accountability of data usage - all the steps for stakeholders are provided; (*iii*) A TDU Ontology (TDUO), created by extending some related concepts of the data usage conceptual model. We also define trust policies based on defeasible rules and perform trust enforcement; and (*iv*) We implement a prototype as a use case based on the TDU architecture to evaluate its performance.

The rest of the paper is organized as follows. Section II presents a motivating scenario to illustrate the need for TDU. Section III presents our proposed system architecture in detail and Section IV discusses the semantics and abstraction. Section V presents the transparency and accountability mechanism. Section VI presents our prototype implementation along with the results. The related work is discussed in Section VII, and finally Section VIII concludes the paper by highlighting few lesson learned and some ideas for future work.

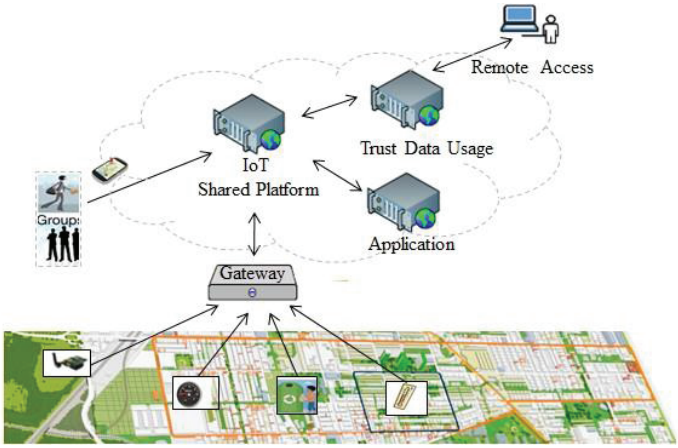

Fig. 1: Motivating Scenario.

## II. MOTIVATING SCENARIO

In this section, we describe a use case scenario to illustrate the need for TDU. First to complete the proposed scenario in our previous work [4], it should be noted that without considering TDU, the end-user applications will not perform well, as it won't be able to offer a better experience.

Figure 1 shows the air pollution monitoring scenario use case in a smart city.

There are multiple stakeholders, such as institutional actors, equipment manufacturers, network operators, infrastructure providers, service providers, and end users, which have a diversity of obligations and constraints in terms of controlling the use of their data. In the scope of this study, we cover four high-level descriptions of the data usage requirements:(*i*) Spatio-temporal granularity; (*ii*) Abstraction/masking of certain information; and (*iii*) the Conditions by class of actor/purpose. The main requirements are explained in the following use case scenario.

1) The data owner (the company that deploys and owns the pollution monitoring sensors) will have full access to all the details generated by all the individual pollution monitoring sensors.

2) For municipal authorities, the data owner is willing to make the average air pollution index per street available on an hourly basis.

3) Only statistical data will be made available to commercial operators, over a specific zone and on a weekly basis.

## III. SYSTEM ARCHITECTURE

This section presents the proposed system architecture, beginning with a description of our previous work on which we based our present solution. A detailed description of different layers and functional entities in the architecture is then presented.

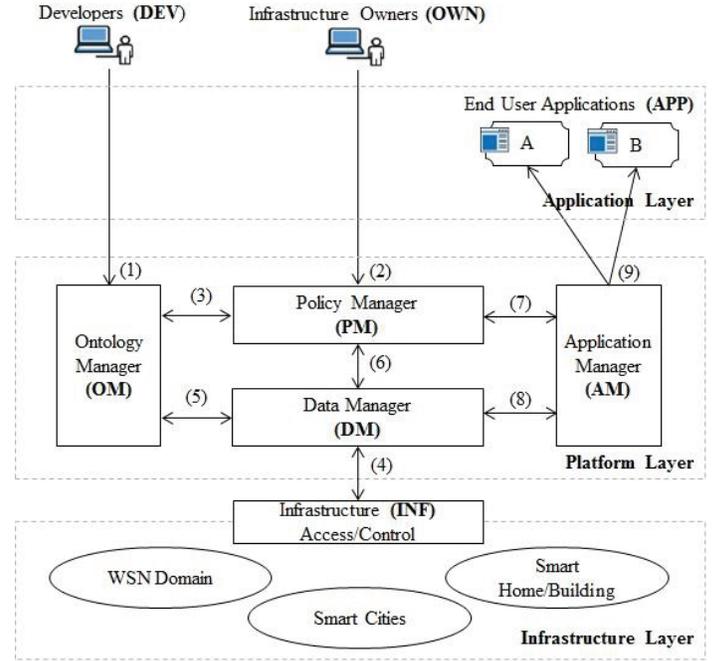

Fig. 2: Proposed TDU Architecture.

### A. Background

The proposed architecture is based on our previous works in [9], [10], and [4]. The architectures in [9] and [10] deal with the simultaneous acquisition of data by multiple applications and services from deployed sensors. These applications and services can be traditional as well as semantic-based Wireless Sensor Network (WSN) applications. When required, the sensor data can be annotated using sensor domain ontology, such as the Semantic Sensor Network (SSN) [11]. However, all of this data is sent directly to the consumers (platform or end-user applications) without allowing the owners of the data to enforce certain policies concerning its usage. In other words, it is assumed that the data is always trusted, which may not be true. For example, issues such as how the same data can be shared among multiple end-users by using different policies based on their location, time or role (home users, city administration or law enforcement agency) are not addressed in the above-mentioned works. In addition, the two architectures mentioned above only consider WSNs as the source of data, whereas in the broader context of the IoT and smart cities, many types of devices, in addition to sensors, provide data to end-users. In [4] we presented a step-by-step data handling mechanism for data owners, data consumers, and an IoT shared platform. These are our staring points for contributing to the architecture proposed in this study.

### B. Layers and Functional Entities

Figure 2 shows the architecture designed for the proposed trust-based data sharing model (TDU). It contains the following three layers:

*1) Infrastructure Layer:* This bottom layer contains a variety of IoT objects that are deployed to send their data to different applications. Because of the IoT scenario, we consider that these IoT objects can belong to different domains, such as smart sensors from the WSN domain, smart street lights/traffic signal poles from smart cities domain, or home alarm systems/intelligent HVAC systems from a smart home/building domain. We also consider that some kind of infrastructure access/control mechanism is used by each of these domains independently.

*2) Platform Layer:* The platform layer is the middle layer, and it consists of the following four functional entities, Ontology Manager (OM), Policy Manager (PM), Data Manager (DM), and Application Manager (AM). This contributes to the advancement of our previous architecture, in which the OM was used to work with the domain and trust ontologies. Here, the PM is used to work with trust policies, the DM is used to work with IoT data or resources from the infrastructure (INF), and the DM works with IoT applications. The interactions between these entities are discussed in Section (V-A).

*3) Application Layer:* The last layer, the application layer, contains end-user applications (APP) that receives the shared data from the infrastructure through the platform. We also consider that in most cases, the APP will receive and consume the sensor data (sent to it according to a pre-set policy) but also the data's owner (OWN) (probably) wants to know the data's usage.

## IV. SEMANTIC AND ABSTRACTION

This section presents the semantic technologies for data integration, modeling, and processing in order to apply this approach to the IoT data in the proposed TDU architecture.

### A. Data integration

In this study, we use semantic technologies to provide data consistency among heterogeneous data set schema. We propose the Resource Description Framework (RDF) [12] to encode the IoT data and resources. Note that the RDF allows for the easy integration of multiple vocabularies [13]. Our IoT data and resources are published as Linked Data [5].

### B. Data modeling

We propose semantic language to model the number of entities and the state of those entities for IoT devices or resources. This makes it possible to interact with higher-level entities rather than directly with IoT devices or resources. Currently, there are numerous efforts to provide ontologies for various domains. For example, for sensors we have an SSN ontology [11] that was developed and proposed at the W3C for standardization. Other ontologies include the Smart Appliance REFerence(SAREF) ontology developed by TNO[1], which covers popular sensors and actuators. Recently, Linked Open Vocabularies for the Internet of Things (LOV4IoT)[2] referenced to more than 300 existing ontology-based projects relevant for the IoT. Introducing abstraction based on a semantic approach is a concept being pushed forward within several standard defining organizations such as the ETSI M2M [14], OneM2M [7], and the W3C Web of Things[3]. Therefore it is worthwhile reusing domain knowledge expertise from the existing ontologies in this architecture.

---
[1]https://www.tno.nl
[2]http://www.sensormeasurement.appspot.com/?p=ontologies
[3]http://www.w3.org/WoT/

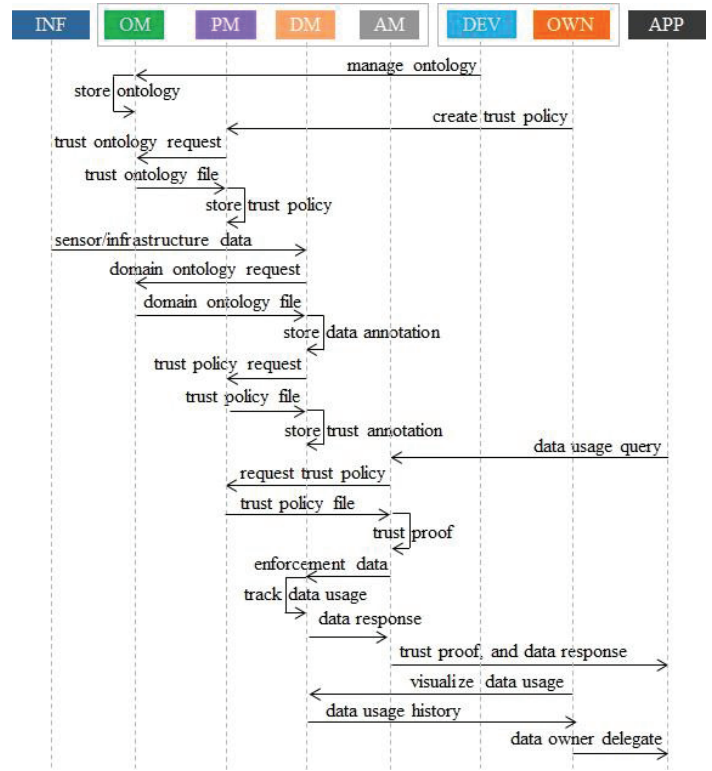

Fig. 3: Transparency and Accountability Mechanism.

### C. Data processing

We use Semantic web technologies to retrieve IoT data by means of SPARQL [15], an SQL-like language that enables querying an RDF store. It also allows logical reasoning, and so new information or knowledge can be inferred from existing assertions and rules. We can re-use existing reasoners (e.g. Pellet[4], Jena[5], or SPINDle [8]) for this purpose.

## V. TRANSPARENCY AND ACCOUNTABILITY

This section presents the mechanism we developed to enhance the transparency and accountability of data usage and illustrate these aspects with sequence diagram. It also includes the trust ontology, trust policy, and reasoning with trust related policies.

### A. Mechanism

Figure 3 shows the sequence diagram of the mechanism to enhance transparency and accountability for data usage. The sequence here is aligned to the steps of the Figure 2 shown by the numbers in the arrows.

The OM is used to manage the trust ontology mentioned in Section (III-B2 and V-B) to provide TDU. This type of ontology can be provided by the developer (DEV) at the platform. Next, trust policies from the data provider, presented in Section (III-B2 and V-C), are managed by the PM. In this study we assume that the owner of the data is the infrastructure owner (OWN). The policies can be obtained by the platform using a simple web-based form. For example, the data owner can specify that the data from its infrastructure can be shared with industrial users but not with academic users. This process can also be used to specify granularities, such as to share data from location A with academic users and to share data from location B with industrial users only. Once the data usage policy is received, the PM creates and stores the trust policy based on the specific rules and the trust ontology. The data from the INF is then sent to the DM and annotated with the metadata to control its usage and/or to make it more trustworthy. The PM extracts the specific rules from the related trust policy for the trust annotation process.

---
[4]https://github.com/complexible/pellet
[5]https://jena.apache.org/

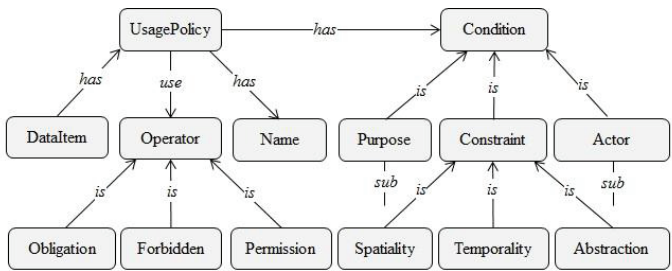

Fig. 4: The proposed Trust Data Usage Ontology (TDUO).

Next, the APP sends a data usage query to the platform; the AM entity is responsible for processing the request. First, it checks for the trust proof with the PM that this query is provable or not. If the request is provable, the DM filters/provides data according to the rules extracted from the policy to the AM. Next, the AM is tasked to keep track of the data usage history from the APP accessing the platform and to send them the final data. Using these steps, it is possible to have a TDU based on the owner-specified policies. The OWN also can request the platform to visualize data usage history, and process the data owner delegation to the APP.

### B. Trust Ontology

This subsection presents the trust ontology used to define the trust policy formulated in the next subsection.

The proposed trust ontology is called TDUO which provides more concepts related to the previously presented data usage conceptual model [4]. We define usage policy by using modal operators (Obligation, Forbidden, and Permission) on the following conditions: (*i*) class of actors, (*ii*) constraints (Spatiality, Temporality, and Abstraction), and (*iii*) class of purposes. The proposed TDUO is illustrated in Figure 4.

*1) Data Items:* A *Data Item* is an individual part of the Entity Element. The *Entity Element* is a container used to exchange information about an entity. It contains the following information: (*i*) an entity ID including the name and the type, (*ii*) a list of the entity attributes, (*iii*) (optionally) the name of an attribute domain that logically groups together a set of entity attributes, and (*iv*) (optionally) a list of metadata that apply to all the attribute values of the given domain. We formally define a *Data Item* by using XML DTD, as mentioned in Listing 1.

```
1  <!DOCTYPE TDUO[
2  <!ELEMENT DataItem(EntityElement)>
3  <!ELEMENT EntityElement(EntityID,
       AttributeDomainName?, EntityAttributeList,
       DomainMetadata?)>
4  <!ELEMENT EntityID(Id, Type)>
5  <!ELEMENT EntityAttributeList(EntityAttribute*)>
6  <!ELEMENT EntityAttribute(Name, Type,
       EntitytValue, EntityMetadata+)>
7  <!ELEMENT DomainMetadata(EntityMetadata*)>
8  <!ELEMENT EntityMetadata(Name, Type, Value)>
9  ---
10 ]>
```

Listing 1: XML DTD Definition of Data Item.

*2) Conditions:* The condition list contains (optionally) the following expressions: (*i*) Temporal Constraints for temporal granularity, (*ii*) Spatial Constraints for spatial granularity, (*iii*) Abstraction Constraints for the masking of certain information, (*iv*) Conditions by Actors, and (*v*) Conditions by Purposes. We formally define conditions by using XML DTD, as shown in Listing 2.

```
1  <!DOCTYPE TDUO[
2  <!ELEMENT Condition(Temporality*, Spatiality*,
       Abstraction*, Actor*, Purpose*)>
3  <!ELEMENT Spatiality(SpatialScope*)>
4  <!ELEMENT Temporality(TemporalScope*)>
5  <!ELEMENT Abstraction(AbstractScope*)>
6  <!ELEMENT Actor(ActorScope*)>
7  <!ELEMENT Purpose(PurposeScope*)>
8  <!ELEMENT TemporalScope(Secondly?, Minutly?,
       Hourly?, Daily?, Weekly?, Monthly?, Yearly?,
       Any?)>
9  <!ELEMENT SpatialScope(Street?, Zone?, Any?)>
10 <!ELEMENT ActorScope(DataOwner?,
       MulnicipalAuthority?, ComercicalOperator?)>
11 <!ELEMENT AbstractScope(Aggregation?, Detail?,
       Any?)>
12 <!ELEMENT PurposeScope(CommercialUse?, Any?)>
13 ---
14 ]>
```

Listing 2: XML DTD Definition of Condition.

*3) Operators:* This is a set of *model operators* (*i*) Obligation (*ii*) Forbidden, and (*iii*) Permission. The formal definition created using XML DTD is presented in Listing 3.

```
1  <!DOCTYPE TDUO[
2  <!ELEMENT Operator(Obligation?, Forbidden?,
       Permission?)>
3  ---
4  ]>
```

Listing 3: XML DTD Definition of Operator.

*4) UsagePolicy:* A collection of rules created by defining *Operators* on the individual *Condition*. Listing 4 formally defines the definition of *Usage Policy* using XML DTD.

```
1  <!DOCTYPE TDUO[
2  <!ELEMENT UsagePolicy(Name,Rule*)>
3  <!ELEMENT Rule(Operator?, Condition?)>
4  <!ELEMENT Name(URI?)>
5  ---
6  ]>
```

Listing 4: XML DTD Definition of Usage Policy.

### C. Trust Policy

The trust policy is used by the stakeholders to define the diversity of obligations and constraints that they wish to impose on the usage of their data in the context of sharing by several smart cities actors. We consider that there are many possible stakeholders' policies depending on the scenarios, sizes of cities, and infrastructures. Since multiple stakeholders each provide their trust policies, this may lead to inconsistent and conflicting policies. To solve the conflicts that will arise between rules and exceptions, we have applied Defeasible Logic (DL) [16] to model the policy [4]. In this section, we describe in detail the trust policies related to our use case scenarios.

Data owners ($DO$): have full access to all the details. This policy is represented with the use of defeasible rules, as follows:

$$R^{DO} = \{r_{1,d} : DO(X) \Rightarrow_P TemporalScope(X, any),$$
$$r_{2,d} : DO(X) \Rightarrow_P SpatialScope(X, any),$$
$$r_{3,d} : DO(X) \Rightarrow_P AbstractScope(X, any),$$
$$r_{4,d} : DO(X) \Rightarrow_P PurposeScope(X, any)\}$$

Municipal authorities ($MA$): have permission to access the available average air pollution index (aggregation), e.g. per street on an hourly basis. This policy is represented with the use of defeasible rules, as follows:

$$R^{MA} = \{r_{1,m} : MA(X) \Rightarrow_P SpatialScope(X, street),$$
$$r_{2,m} : MA(X) \Rightarrow_P TemporalScope(X, hourly),$$
$$r_{3,m} : MA(X) \Rightarrow_P AbstractScope(X, aggregation)\}$$

Commercial operators ($CO$): only statistical data will be made available, e.g. over a zone and on a weekly basis. This policy is represented with the use of defeasible rules, as follows:

$$R^{CO} = \{r_{1,c} : CO(X) \Rightarrow_P SpatialScope(X, zone),$$
$$r_{2,c} : CO(X) \Rightarrow_P TemporalScope(X, weekly),$$
$$r_{3,c} : CO(X) \Rightarrow_P AbstractScope(X, statistic)\}$$

### D. Trust Enforcement

For example, we may have a consumer's request that a commercial operator ($CO$) requests all the details of the air pollution index data over a street on an hourly basis. This request is represented with the use of defeasible rules, as follows:

$$r : CO(X), [P]SpatialScope(X, street),$$
$$[P]TemporalScope(X, hourly),$$
$$[P]AbstractionScope(X, detail)$$
$$\Rightarrow_O ConsumerRequest(X)$$

Reasoning with the trust policy, we come to the conclusions: $-\Delta[O]ConsumerRequest(X)$, $-\partial[O]ConsumerRequest(X)$. This means that this $ConsumerRequest$ is not defeasible provable, and so the request is refused. If the consumer's request is defeasible provable, the related data items will be filtered or aggregated following the request conditions before returning the results to the consumer. Every data usage transaction will be recorded as a new data item and later reported to the data owners.

## VI. PROTOTYPE IMPLEMENTATION AND RESULTS

To validate the proposed solution in this study, we implemented a prototype illustrating an air pollution monitoring use case and conducted some performance analysis.

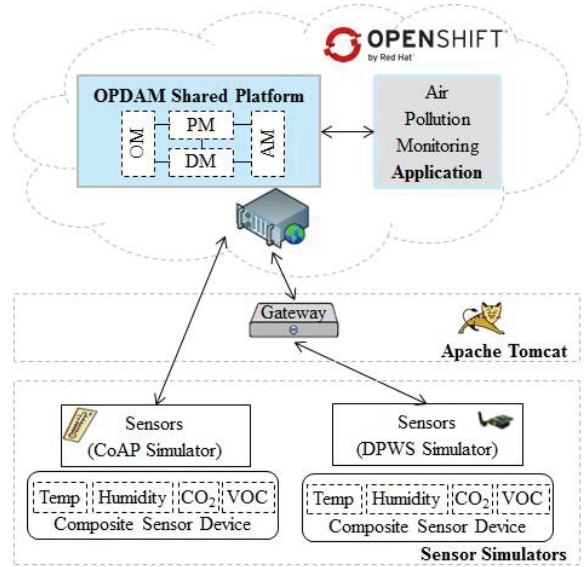

Fig. 5: Implemented Architecture of the Prototype.

### A. Implementation Consideration

We developed an air pollution monitoring application (APM) based on the scenario presented in Section II. It is offered as a RESTful web service based on Java Technology. The application was deployed in a cloud-based OPENSHIFT[6], which is a Platform as a Service (PaaS) that allows the development of SaaS applications without having to maintain a server. Figure 5 presents our implemented architecture for the prototype.

*1) Sensors:* We simulated two composite sensor devices by using DPWS Simulator[7], and CoAP Simulator[8]. These sensors are used to measure air pollution indexes such as temperature, humidity, CO2, and VOC (Volatile organic compound) data.

*2) Gateways:* We used Apache Tomcat[9] to deploy a web application server for a gateway simulator. It received data from the sensors.

*3) OPDAM Platform:* We used Apache Jena Framework[10], an open source Java Framework for developing semantic technology. We also used SPINdle [8], a logic reasoner that can be used to compute the consequence of DL theories in an efficient manner. We built our shared platform, called OPDAM, including four main components: OM working with Air Pollution Domain Ontology and Trust Ontology; PM with Trust Policy/Rules; DM with sensor data, data annotation, and trust annotation; and DM with data enforcement and track data usage.

*4) APM Application:* The APM is a RESTful service developed using Restlet[11], a framework for developing REST web services. The service requests the relevant air pollution data from the OPDAM platform.

---

[6]https://www.openshift.com/
[7]https://github.com/sonhan/dpwsim
[8]https://github.com/caohuuquyet/jhess/tree/master/jUCP
[9]http://tomcat.apache.org/
[10]https://jena.apache.org/
[11]http://restlet.com/

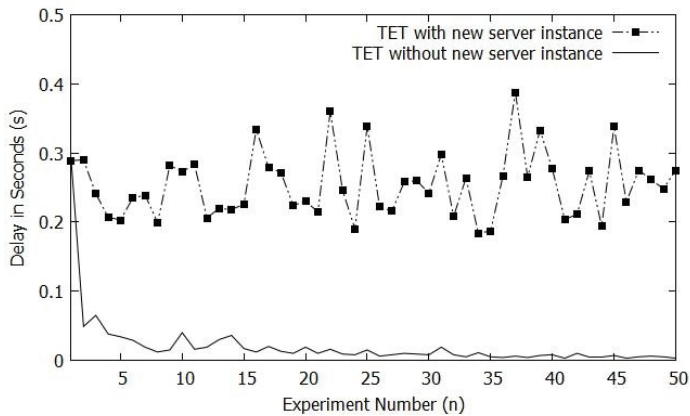

Fig. 6: Trust Enforcement Time.

*B. Prototype Setup*

We set up the prototype based on the architecture shown in Figure 5. First, we deployed two virtual sensors developed in Section (VI-A1). We also simulated sensors in different virtual areas to measure air pollution in those areas. One sensor sent its data to the gateway simulator outlined in Section (VI-A2), and then the data was transferred to the platform by the gateway. Other sensors sent data directly to the platform. The OPDAM platform developed in Section (VI-A3) and the APM service developed in Section (VI-A4), are deployed on the cloud[12].

*C. Preliminary Performance Analysis*

This part mainly presents the results based on tests for the trust enforcement time (TET) in the implemented architecture. The goal is to show how much overhead is incurred due to the enforcement of data usage control based on the implementation configuration in Section VI-B.

To evaluate the proposed solution, we performed two experiments and repeated each experiment 50 times. Their confidence interval is 95%. In the first experiment, the server was restarted to create new server instance for each trust enforcement request. In the second experiment, we used the same instance of server for subsequent trust enforcement requests. Based on that, we measured the TET in delay seconds. Figure
6 shows the results of each experiment (with and without new server instance respectively).

If server is restarted (new server instance is created for each request) for each trust enforcement request we had average delay of over 253ms. If same instance of server is used for subsequent trust enforcement requests we have very less delay, the average time was around 20ms. In most cases we did not have large delay, hence trust enforcement does not incur much additional delay.

We also tried to use a cloud-based Google App Engine (GAE) but encountered exceptions such as access control exception while invoking the SPINdle logic reasoner. Solving these exceptions involves deep understanding of GAE and improve other performance analysis aspects, this is our planned future work.

## VII. RELATED WORK

Several research activities have investigated supporting for confidence related to data sharing in different domains, such as Web, Social Networks, Ubiquitous Computing, WSN, and IoT/Smart Cities. We categorized the different axes of confidence as follows: (*i*) Traditional security mechanisms (Access control, Authorization, Accountability [17], Privacy [18] [19]); (*ii*) Precision, Reliability and Trust Issues [20] [21]; (*iii*) Abstraction and masking of information (for example, which level of information should be shared) [13]; (*iv*) Data Licensing [22]; and (*v*) Usage control (how data is used after access to it has been granted [23]) and usage control mechanisms are well studied and continue to be improved [24]. In this study, we proposed a trusted data usage approach in the context of a shared platform in smart cities. We did not focus on security aspects such as confidentiality, access control, or privacy. In fact, we used the concept of usage control [23] as a starting point and then the data usage conceptual model [4] to propose our trust data usage concepts by defining the data usage requirements based on spatio-temporal granularity, the abstraction/masking of certain information and conditions by class of actors or purposes. We then focused our study on the architecture to ensure data usage transparency and accountability. To the best of our knowledge, the ideas proposed in this study are novel and different from the previous efforts in the literature.

## VIII. CONCLUSION AND FUTURE WORK

Trust is the key for sharing IoT data among various stakeholders. Using a simple scenario for smart cities, we propose a trust model to harmonize data sharing incorporating policies defined by the data owner. In summary, we have contributed a novel multi-layer architecture for TDU including a use case scenario, its background, main functional entities, and semantic and abstraction models. The mechanism for transparency and accountability of data usage has provided as a sequence diagram to the smart cities' stakeholders. This also has proposed a TDUO trust ontology, defined trust policies based on the trust ontology and defeasible rules, and performed trust enforcement using defeasible reasoning. We finally have implemented a prototype based on the TDU architecture to evaluate our solution.

We learned several lessons from this study. The first is that multiple stakeholders have to be involved in defining data sharing policies. In simple scenarios, these are normally defined by the data producer. However, in the IoT, data is often merged from various sources and it becomes difficult to determine who owns which data. The second lesson concerns the trust ontologies. We found that some effort needs to be expended in defining general and more open trust ontology, a likely topic for future work. The third lesson is that the defining of trust and data usage policies should have input from multiple actors. How data from IoT devices in a private domain (e.g. smart homes) can be utilized and/or provided to interested entities needs to be explored. In some cases giving incentives, like tax rebates, will be useful.

For the next steps in this work, we will consider its implementation and validation in the real environment using end-to-end interactions. Another future work item is the development of an open source rule engine interpreter. This rule engine can be used on the cloud-based GAE with no encountered exceptions and improved performance aspects. Defeasible logic is a useful technique to check for inconsistent rules, but we need to explore if there are other solutions available for this purpose. Lastly, we are going to provide a visualization tool to help users to customize their policies in an interactive format that allows them to explore the consequences of certain changes.

---

[12]http://opdam2-opdam.rhcloud.com/

ACKNOWLEDGMENT

The research leading to these results was partially funded by the ITEA projects Fuse-IT, SEAS and CAP, and the ICT R&D program of Information & Communications Technology Promotion (IITP) funded by the Korea government (MSIP) [R0190-15-2027, Development of TII (Trusted Information Infrastructure) S/W Framework for Realizing Trustworthy IoT Eco-system].